\documentclass[
]{llncs}

\newenvironment{mytable}{
\vspace{0.2ex}
\begin{center}\begin{minipage}
{0.9\textwidth}\renewcommand{\baselinestretch}{0.75}\begin{small}}
{\end{small}\end{minipage}\end{center}
\vspace{0.1ex}
}
\newcommand{\mycaption}[1]{\refstepcounter{table}\centerline{\textbf{Table~\arabic{table}.} #1}}
\newcommand{\afp}{AFP}
\newcommand{\att}{ATT}
\newcommand{\fee}{fee}
\newcommand{\he}{she}

\newcommand{\his}{her}

\newcommand{\himself}{herself}
\newcommand{\junction}{compatibility}
\newcommand{\Junction}{\xmakefirstuc{\junction}}
\newcommand{\hol}{higher order logic}
\newcommand{\Hol}{HOL}
\newcommand{\I}{Isabelle}
\newcommand{\fo}{ForMaRE}
\newcommand{\M}{Mizar}
\newcommand{\query}[1]{\marginnote{\raggedright\footnotesize\itshape\hrule\smallskip{#1}\smallskip\hrule}}
\renewcommand{\query}[1]{} 
\newcommand{\rnote}[1]{\query{#1}}
\newcommand{\lnote}[1]{\reversemarginpar\query{#1}\normalmarginpar}

\usepackage{graphicx}
\usepackage{wrapfig}
\usepackage{alltt}
\usepackage{fancyvrb}
\usepackage[T1]{fontenc}
\usepackage[utf8]{inputenc}
\usepackage{setspace}
\usepackage{unicode-chars}

\usepackage{amsmath} 
\usepackage{amssymb}
\usepackage[]{marginnote}
\usepackage[]{mfirstuc}

\usepackage[british]{babel}
\usepackage[babel]{csquotes}
\MakeAutoQuote{“}{”}
\MakeAutoQuote*{‘}{’}

\usepackage[natbib=true
, firstinits=true
, isbn=false
, url=false
, backend=bibtex
, maxbibnames=2
, maxcitenames=3]{biblatex} 
\begin{document}

\title{Set Theory or Higher Order Logic to Represent Auction Concepts in Isabelle?\thanks{This work has been supported by EPSRC grant EP/J007498/1 and an LMS Computer Science Small Grant.
\dots The final publication is available at http://link.springer.com.
}}


\author{Marco B.\ Caminati\inst{1}
\and Manfred Kerber\inst{1}
\and Christoph Lange\inst{1,2}
\and Colin Rowat\inst{3}
}

\institute{%
Computer Science, University of Birmingham, UK%
\and Fraunhofer IAIS and University of Bonn, Germany%
\and Economics, University of Birmingham, UK\\
Project homepage: \url{http://cs.bham.ac.uk/research/projects/formare/}
}

\maketitle
\lnote{MC: Main revisions implementing reviewers comments have a sidenote like this one. In the \LaTeX{} source, I appended a checklist of TODO changes, which I'm following.}
\begin{abstract}
When faced with the question of how to represent properties in a formal proof system any user  has to make design
decisions. We have proved three of the theorems from Maskin's 2004
survey article on Auction Theory using the \I/HOL system, and we have verified software code
that implements combinatorial Vickrey auctions.
A fundamental question in this was how to represent some basic concepts:
since set theory is available inside \I/HOL, when introducing new definitions there is often the issue of balancing the amount of
set-theoretical objects and of objects expressed using entities which are more typical of higher order logic such as functions or lists.
Likewise, a user has often to answer the question whether to use a constructive or a non-constructive definition. 
Such decisions have consequences for the proof development and the usability of the formalization. 
For instance, sets are usually closer to the representation that economists would use and recognize, while the other objects are closer to the extraction of computational content. 
We have studied the advantages and disadvantages of these approaches, and their relationship, in the concrete application setting of auction theory.
In addition, we present the corresponding \I{} library of definitions and theorems, most prominently those dealing with relations and quotients.
\end{abstract}



\section{Introduction} \label{sec:introduction}
\label{RefIntro}
\lnote{MC: Rephrased after R3 noticed there is also type theory as a third strong competitor.}
When representing mathematics in formal proof systems, alternative foundations can be used, with two important examples being set theory (e.g., \M{} takes this approach) and higher order logic (e.g., as in Isabelle/HOL). 
Another dimension in the representation is the difference between classical and constructive approaches. 
Again, there are systems which are predominantly classical (as most first order automated theorem provers) and constructive (e.g., Coq). 
\I{}/HOL is flexible enough to enable the user to take these different approaches in the same system (e.g., although it is built on higher order logic, it contains a library for set theory, \verb|Set.thy|).
For instance, participants in an auction, i.e.\ \emph{bidders}, can be represented by a predicate \verb|Bidder x| or alternatively as a list of bidders \verb|[b1, b2, b3]|.

The difference between a classical and a constructive definition can be demonstrated by the example of 
\lnote{MC: Changed `maximal argument' to `argument of the maximum' for Reviewer2.}
the argument of the maximum for a function (which we need for determining the winner of an auction).
Classically we can define it, e.g., as 
\Verb[commandchars=\\\{\}]|arg_max f A = \{x \(\in\) A. f x = Max(f`A)\}|. 
This definition is easy to understand but, unlike a constructive one, it does not tell how to compute \verb|arg_max|. 
The constructive definition is more complicated (and has to consider different cases). 
It corresponds to a recursive function recurring on the elements of the domain.
From a programming perspective, the two kinds of definitions just illustrated (classical versus constructive) can be seen respectively as \emph{specification} versus \emph{implementation}.



The approaches coexist in \I{}/HOL. 
For example, for a set \verb|X| one can apply the higher order function \verb|f| by the construction \verb|f ` X| to yield the image of the set. 
As a result, an author does not have to make a global decision of whether to use sets or higher order functions, but has fine grained control on what to use for a new mathematical object (e.g., sets vs.\ lists or lambda functions).  
Such a choice will typically depend on many factors. 
One factor is the task at hand (e.g., whether one needs to prove a theorem about an object or needs to compute its value). Another factor is naturalness of the constructions. 
This will typically depend on the authors and their expected audiences.
Pragmatically, users also have to consider which of the possible approaches is more viable given the support provided by existing libraries for the system being used.


The \fo{} project~\cite{LRK:FormareProject13:short} applies formal methods to economics.
\lnote{MC: Added this sentence for reviewer 1.}
One of the branches of economics the project focuses on is \emph{auction theory}, which deals with the problem of allocating a set of resources among a set of participants while maximizing one or more parameters (e.g., revenue, or social welfare) in the process.
\fo{} has produced the Auction Theory Toolbox (\att{}), containing \I{} code for a range of auctions and theorems about them.%
\footnote{See \url{https://github.com/formare/auctions/}; 
the state at the time of this writing is archived at \url{https://github.com/formare/auctions/tree/1f1e7035da2543a0645b9c44a5276229a0aeb478}.
}  
Therefore, there is a good opportunity to practically test the feasibility of the different approaches, as introduced above, in this concrete setting.
We adopted a pragmatic attitude: we typically took a set theoretic approach, since firstly we felt most familiar with it and secondly we knew from our ongoing interaction with economists that it would look more natural to them. However, when we needed to generate code, this generally excluded set-theoretical constructions such as the set comprehension notation.
As a consequence most of our work was done in the set-theoretical realm, but some done constructively which allowed us to produce the code we wanted. 
The disadvantage of this approach is that we had to provide supplementary `\junction{}' proofs to show the equivalence of the set theoretical and the computable definitions when needed.
This means that, as a byproduct of our efforts, we also generated a good amount of generic set-theoretical material which was neither provided by the \I{} library nor \I{}'s Archive of Formal Proofs (\afp{})\cite{afp}.

In this paper, rather than discussing our progress in the application domain (auction theory), we illustrate this material and its relationship to the mathematical objects we needed.
We will discuss this by the following three concrete examples that occurred while we were working on the ATT:
\begin{enumerate}
\item
We model a function indirectly through the set-theoretical notion of the graph of a relation, rather than directly using the HOL primitive notion of a lambda-abstracted function. This allowed us to concisely define auction-related notions through two natural set-theoretical constructions (extension and restriction of a function). Moreover, this in turn allowed us to generalize some theorems from functions to relations. 
We also discuss how this choice does not necessarily mean giving up the advantage of computability (section \ref{RefFuncAsGraph}).
Incidentally, this will also give insight into the main differences between set theory as implemented in \I{}/HOL and standard set theory. 
\item
We developed a stand-alone formalization of the definition of functions over equivalence classes (section \ref{RefQuotients}). 
This is a common construction when, e.g., defining the canonically induced operation on the quotient of a group by one of its normal subgroups.
To define such an operation, we proved the invariance (or well-definedness) under an equivalence relation.
\item
\label{RefDualDefsItem}
We applied a set-theoretical, non-constructive definition and a constructive one to the same object (the set of all possible partitions of a set). The two approaches were used for different purposes (theorem proving and computation, respectively) and proved formally their equivalence by a 
`\junction' theorem (section \ref{RefDualDefs}).
The same approach is adopted for the set of all possible injective functions between sets.
\end{enumerate}
In existing proof assistants based on set theory (e.g., \M{}, Metamath), the desirable quality of being able to compute values of functions (rather than only the truth value of predicates involving them), is typically lost%
\lnote{MC: Added footnote for reviewer3.}%
\footnote{We do not know if this is due to how the existing proof assistants are implemented, or to some fundamental limitation of set theory. 
Reading the recent ``Computational set theory{}'' thread (\url{http://www.cs.nyu.edu/pipermail/fom/2014-February/017841.html}), and its related threads, on the ``Foundations of Mathematics'' mailing list, we feel the question is currently 
open.}%
.

We will present examples which show that we can retain this \I{}/HOL-induced advantage in our developments.

\subsection*{Overview of the paper}
Section \ref{RefFuncAsGraph} introduces and motivates the set-theoretical encoding of functions, as an alternative to a lambda representation; this is done by commenting on the relevant \I{} definitions, and by illustrating a particular theorem regarding auctions, which employs those definitions.
Section \ref{RefQuotients} brings this approach to a more abstract level, by showing how it can handle, given an initial function, the definition of a second function on the corresponding equivalence classes. Since this was already done in \I{}'s \verb|Equiv_Relations.thy|, we also illustrate the differences with it.

In some cases, defining a mathematical object in purely set-theoretical terms does not preserve computability of that object: section \ref{RefDualDefs} introduces a technique for such cases%
.
As mentioned in point \ref{RefDualDefsItem} above, this technique introduces a parallel, computable definition, and then formally proves the equivalence of the two definitions.
In the same section, this technique is presented through two examples from our \att{}; we then discuss how we took advantage from having preserved both the definitions.
Finally, section \ref{RefApplication} explains how we applied the general machinery from section \ref{RefQuotients} to the \att{}.


\section{Set-theoretical Definition of Functions in \I{}/HOL}
\label{RefFuncAsGraph}
In \hol{}, functions, function abstraction, and function application are primitives~\cite{bowen1994z};
in set theory, the primitive notion is that of a set, and a function $f$ is represented by its graph, i.e., the \emph{set} of all the ordered pairs $\left( x, f\left( x \right) \right)$.
We chose to work mostly with the set-theoretical representation of the functions even though we are using a tool based on \hol{}. In the following we give reasons why.
\begin{enumerate}
\item
\label{RefMotivationEnumeration}
A first reason is that a set theoretical representation more easily allows to enumerate all the (injective) functions from a finite set to another finite set (\cite{CKLR:SoundCombVickCode13}).
In contrast, this seems to be more complicated to do directly in \hol, where all the functions are assumed be total.
\item
\label{RefMotivationRelations}
A second reason is that the set-theoretical, graph-based representation works even for generic relations, thus often allowing us to extend the results we proved to the more general situations involving relations, rather than functions (see the next subsection).
\item
\label{RefMotivationRestriction}
A third reason is that the operation of function restriction is naturally expressed in terms of two elementary set operations: cartesian product and intersection. Indeed, this allows to extend this operation to relations, immediately giving an instance of what we argued in point~\eqref{RefMotivationRelations} above.
Restriction is a fundamental operation for representing the concept of weakly dominant strategy, a key concept in auction theory, see e.g., \cite[proposition~2]{mas-04}.
The definition of function restriction is arguably more complicated in standard \hol{} since functions are always total. That is, restricting them to a set requires carrying the restricted domain set with the function, e.g.\ by forming a pair $(R, f|_R)$, whereas the set-theoretical representation naturally includes this set.
\item
Finally, specific partial, finite set theoretical functions can be very concisely and quickly defined; this made it possible to  prompt\-ly test \I{} code while we were working on it.
For example, functions can be written in the form of a set as \verb|{(0,10), (1,11), (2,12)}|  and fed to the \I{} definitions very easily in order to test the correctness of related computations empirically, whereas a lambda expression would be more complex to define.\rnote{MK: Added ``to define''}
\end{enumerate}

\subsection{Two basic operators on relations: “outside” and pasting}
\label{RefOutside}
In this section, we discuss two general mathematical operations we encountered specifically during formalization of auctions. Assume a set of bidders 
$N
$ and a function $b:N\to\mathbb{R}$ that determines the corresponding bids.
The first operation removes one bidder $i$ from the domain $N$ of the bid function.
The second operation alters the bid function in one point $i$ of its domain, $b\,+\!*\,(i,b_i)$, which is equal to $b$ except for argument $i$, where the value is changed to $b_i$.
In set theory, a function (or a relation) inherently specifies its domain and range.
A generalization of the first operation is thus obtained by writing (in \I):
\begin{mytable}
\begin{alltt}
definition Outside :: 
"('a \( \times \)  'b) set ⇒ 'a set ⇒ ('a × 'b) set" 
(infix "outside" 75) where 
"R outside X = R - (X × Range R)",
\end{alltt}
\end{mytable}
where 75 denotes the binding strength of the infix operator. The following specialization to singletons …
\begin{mytable}
\begin{alltt}
abbreviation singleoutside (infix "--" 75) where 
"b -- i ≡ b outside \{i\}",
\end{alltt}
\end{mytable}
… turned out handy for our purposes.

Another circumstance making the set theoretical approach convenient is that now the second operation can be obtained in simple terms of the first as follows:
\begin{mytable}
\begin{alltt}
definition paste (infix "+*" 75)
where "P +* Q = (P outside Domain Q) ∪ Q",
\end{alltt}
\end{mytable}
which can be specialized to the important case of \verb|Q| being a singleton function:
\begin{mytable}
\begin{alltt}
abbreviation singlepaste where 
"singlepaste F f ≡ F +* \{(fst f, snd f)\}"
notation singlepaste (infix "+<" 75).
\end{alltt}
\end{mytable}

While we often applied \verb|outside| and \verb|+*| to functions rather than relations (i.e., assuming right uniqueness, see section~\ref{RefRuniq}), many of its properties were proved for relations in general.  For example, the associativity theorem
\begin{mytable}
\begin{alltt}
lemma ll53: "(P +* Q) +* R = P +* (Q +* R)"
\end{alltt}
\end{mytable}
has no hypotheses in its statement and holds for general relations.

\subsection{Specializing relations to functions: right-uniqueness and evaluation}
\label{RefRuniq}
The operators \verb|outside|
and \verb|+*| are 
building blocks on which the statements of many theorems we proved for auctions are based.
In turn, a number of preparatory lemmas have been proven about those objects in the generic case of relations; however, others only hold when considering relations which actually are functions.
Hence, we need the following predicate for \emph{right-uniqueness} of a relation, which we define in terms of the \emph{triviality} of a set (i.e.\ being empty or singleton):
\begin{mytable}
\begin{alltt}
definition trivial where "trivial x = (x ⊆ \{the_elem x\})",
\end{alltt}
\end{mytable}
where \verb|the_elem| extracts an element from a set, being undefined when this cannot be done in a unique way.\rnote{MK: changed `one unique way' to `a unique way'}
The predicate for right-uniqueness is called \verb|runiq|, and it uses the operator \verb|R``X|, which yields the image of the set \verb|X| through the relation \verb|R|:
\begin{mytable}
\begin{alltt}
definition runiq :: "('a × 'b) set ⇒ bool" where
"runiq R = (∀ X . trivial X → trivial (R `` X))".
\end{alltt}
\end{mytable}

We note that, contrary to other proof assistants based on set-theory (e.g., \M{}), which in general cannot directly compute values of functions (at most the truth or falsity of predicates involving those values), we are able to preserve from \I{}/HOL the ability of actually computing the evaluation of these set-theoretical flavoured functions, when they are right-unique, through the following operator:
\begin{mytable}
\begin{alltt}
fun eval_rel :: "('a × 'b) set ⇒ 'a ⇒ 'b" (infix ",," 75)
where "R ,, a = the_elem (R `` \{a\})"
\end{alltt}
\end{mytable}
Now, indeed, set theoretical functions can be evaluated via \verb|,,|.%
This can be tested through the \I{} command \verb|value|: for example we can write
\begin{mytable}
\begin{alltt}
value "\{(0::nat,10),(1,11),(1,12::nat)\} ,, 0"
\end{alltt}
\end{mytable}
and obtain \verb!10! as an answer.
This holds also when combining \verb|eval_rel| with the operators as from the beginning of this section; for example
\begin{mytable}
\begin{alltt}
value "(\{(0::nat,10),(1,11),(1,12)\} +< (1,13::nat)) ,, 1"
\end{alltt}
\end{mytable}
yields the answer \verb|13|.\rnote{MK@MC: should $\tt +<$ not be $\tt +*$? And on page 9 as well?}

A right-unique relation and a standard \hol{}, lambda abstracted function represent the same mathematical object, hence it should be possible to pass from one representation to another. 
\verb|graph| from \verb|Function_Order.thy|, defined as
\begin{mytable}
\begin{alltt}
definition graph where "graph X f = \{(x, f x) | x. x ∈ X\}" 
\end{alltt}
\end{mytable}
does exactly that. 
The opposite conversion can be achieved easily as follows:
\begin{mytable}
\begin{alltt}
definition toFunction (* inverts graph *)
where "toFunction R = (λ x . (R ,, x))"
\end{alltt}
\end{mytable}

However, the degree of computability of set-theoretical functions is less than with original HOL functions; for example we cannot evaluate
\begin{small}
\begin{equation*}
\label{RefFunctionNotEvaluable}
\verb|value "(graph {x::nat. x<3} (λx. (10::nat))),,(1::nat)"|,
\end{equation*}
\end{small}

\noindent while the following works as expected
\begin{mytable}
\begin{alltt}
value "(graph \{0,1,2\} (λx. (10::nat))),,(1::nat)".
\end{alltt}
\end{mytable}

We also note that, since Isabelle formalizes set theory inside \hol{}, types still impose some rigidity, compared to stand-alone set theory: see section~\ref{RefSetTheory}.
For example, the following alternative \I{} definition would be exactly equivalent to \verb|eval_rel| in a standard (untyped) set theory:
\label{RefEvalRel2}
\begin{mytable}
\begin{alltt}
abbreviation "eval_rel2 (R::('a×('b set)) set) (x::'a) 
≡ ⋃ (R``\{x\})" notation eval_rel2 (infix ",,," 75),
\end{alltt}
\end{mytable}
It is, however, actually defined (and equivalent to it) \emph{only} for set-yielding relations:\rnote{MK: split long sentence into two.}
\begin{mytable}
\begin{alltt}
lemma lll82: assumes "runiq (f::(('a × ('b set)) set))"
"(x::'a) ∈ Domain f" shows "f,,x = f,,,x"
\end{alltt}
\end{mytable}
However, when it is applicable, \verb|eval_rel2| has the desirable qualities of evaluating to the empty set outside the domain of \verb|f|, and in general to something defined when right-uniqueness does not hold (in which case \verb|eval_rel| is undefined).
This allowed us to give more concise proofs in such cases.


\verb|runiq| is a central definition in our formalization; its many possible equivalent formulations have 
turned out to be useful in different steps when proving various lemmas.
Here we present the possible alternative definitions we have proven to be equivalent in the \att{}: 

\begin{mytable}\renewcommand{\baselinestretch}{0.65}
\begin{alltt}
lemma lll33: "runiq P=inj_on fst P"

lemma runiq_alt: "runiq R ↔ (∀ x . trivial (R `` \{x\}))"

lemma runiq_basic: 
"runiq R ⟷ (∀ x y y' . (x, y) ∈ R ∧ (x, y') ∈ R ⟶ y = y')"

lemma runiq_wrt_eval_rel: 
"runiq R ⟷ (∀x . R `` \{x\} ⊆ \{R ,, x\})"

lemma runiq_wrt_eval_rel': 
"runiq R ⟷ (∀x ∈ Domain R . R `` \{x\} = \{R ,, x\})"

lemma runiq_wrt_ex1: 
"runiq R ⟷ (∀ a ∈ Domain R . ∃! b . (a, b) ∈ R)"

lemma runiq_wrt_THE: 
"runiq R ⟷ (∀ a b . (a, b) ∈ R ⟶ b = (THE b . (a, b) ∈ R))"
\end{alltt}
\end{mytable}

In general, we found that, especially for basic and ubiquitous concepts such as \verb|runiq|\rnote{MK: put runiq into verbose and `as' to `such as'}, the more equivalent definitions we have\lnote{MK: `one has' to `we have'}, the better.
One reason is that this improves the understandability of the formalization: different readers will find different definitions easier to grasp.
Another reason is that automated theorem proving tools, such as Sledgehammer%
\footnote{Sledgehammer is an \I{} tool that applies automatic theorem provers (ATPs) and satisfiability-modulo-theories (SMT) solvers to automatically produce proofs~\cite{isabelle-sledgehammer}.}%
, will be more likely to find automated justification in single steps of subsequent proofs: by picking the appropriate equivalent definition, sledgehammer can find a justification, while, upon removing that definition, it is no longer able to do that. 
We actually experienced this phenomenon with proofs involving \verb|runiq|:
%
the form of \verb|runiq| given in \verb|lll33| 
allowed Sledgehammer to find the proof of this technical lemma:

\begin{mytable}
\begin{alltt}
lemma lll34: assumes "runiq P" shows "card (Domain P) = card P".
\end{alltt}
\end{mytable}

\noindent \verb|lemma lll34| above was in turn used to formalize proposition 3 from \cite{mas-04}.

\subsection{Application to auctions}
\label{RefFuncAsGraphApplication}
Next, we give one example of the roles of the operations introduced in this section
in our practical setting of auctions, for the simple case of a single-good auction.
In this case, the input data for the auction are given through a function \verb|b| associating to each bidder the amount \he{} bids for the good.
Given a fixed bidder \verb|i|, the outcome of the auction is determined by two functions, \verb|a| and \verb|p|.
Both take \verb|b| as an argument: the first yields whether that bidder won the item (\verb|a,,b = 1|) or not (\verb|a,,b = 0|); the second, \verb|p,,b|, yields how much \he{} has to pay.
We take the \I{} formalization of the second proposition in \cite{mas-04}, which is theorem \verb|th10| in file \verb|Maskin2.thy|.
This result proves, given some general requirements, the logical equivalence of two properties, each binding \verb|a| and \verb|p|:
\begin{itemize}
\item
The first property we called \verb|genvick| (for generalized Vickrey auction, see below); it states that the payment imposed to \verb|i| is the sum of a `\fee{}' term \verb|t(b--i)|, 
to be paid irrespectively of the outcome, and of a proper price term \verb|(a,,b - a1)*w(b--i)|, 
which is to be added only in case \he{} obtains the good.
Moreover, the first term does not depend on how much \verb|i| \himself{} bids, but only on others' bids.
The proper price is determined by the auxiliary function \verb|w|, which also does not depend on \verb|i|'s bid (since Vickrey auctions are second price auctions).
Hence, \verb|i|'s bid can influence only whether \verb|i| pays or not the proper price, but not its amount.
\item
The second property, called \verb|dom4|, states that \verb|i| can never be worse off if \he{} changes \his{} bid to \his{} real valuation \verb|v| of the good.
\end{itemize}
Hence, \verb|genvick| assumes the following form:
\begin{mytable}
\begin{alltt}
abbreviation genvick where "genvick a p i w ≡ 
(∃ (a1::allocation) t. (∀ b ∈ Domain a ∩ (Domain p). 
p,,b  =  (a,,b - a1)*w(b--i) + t(b--i)  ))",
\end{alltt}
\end{mytable}

\noindent while \verb|dom4| is the following inequality:

\begin{mytable}
\begin{alltt}
definition dom4 where "dom4 i a p = (∀ b::bid. ∀ v.(
\{b,b+<(i,v)\}⊆(Domain a ∩ (Domain p)) ∧ i∈Domain b)⟶ 
v*(a,,b)-(p,,b) ≤ v*(a,,(b<(i,v)))-(p,,(b+<(i,v))))".
\end{alltt}
\end{mytable}

The operators \verb|--| and \verb|+<| are central in expressing those two conditions; 
their respective general properties (collected in \verb|RelationProperties.thy|) permitted to streamline the proof of the theorem, whose thesis reads:
\begin{equation}
\label{RefMaskin2Thesis}
\verb|genvick a p i w = dom4 i a p|.
\end{equation}
\rnote{MK: Can we formulate the theorem also in plain English?}

\section{Quotients between relations}
\label{RefQuotients}
We built a library of basic facts centred around our new constructs of right-uniqueness (\verb|runiq|), evaluation (\verb|,,|), pasting (\verb|+*|), 
and considering a function \verb|outside| some subset of its domain.
The library also contains more advanced results.
In particular, we describe here our approach to building quotients; then, we derive functions on equivalence classes of points from functions defined on single points.
These methods are common in many areas of mathematics, especially algebra and topology; they are used when a given property holds on classes of objects, and one wants to abstract away from the specific representative of a class, and rather define the given property on the whole class. 
Such classes are typically the equivalence classes induced by an equivalence relation over its domain, which form the \emph{quotient} of the original set.

For example, in group theory, the operation of a group $G$ is canonically transported to the set of the cosets yielded by a normal subgroup $N$. 
This is what makes the quotient group $G/N$ a group, and is only possible if the group operation is \emph{class-invariant}%
\footnote{In this case we can also say that the group operation is \emph{well-defined}, or that it \emph{respects} the corresponding equivalence relation, or even that it is \emph{compatible} with it.}%
: 
the product of two representative elements of two cosets must be in the same coset, irrespective of how those representative elements are selected. 
This is ensured by the definition of normal subgroup.

In our case, to construct \verb|t| 
appearing in the definition of \verb|genvick| towards the end of the previous section, 
we need to define some function taking as an argument a bid vector with the $i$-th component removed (where $i$ is a bidder).
The formal way to do that was to aggregate all possible bid vectors differing only on their $i$-th component and to define that function on the classes obtained this way; hence, using quotients naturally emerged as one elegant approach.
More details on this particular application of quotients are in section \ref{RefApplication}.

As illustrated in \cite{paulson2006defining}, existing \I{}'s theory \verb|Equiv_Relations.thy| already introduces tools for these general mathematical techniques.
There are, however, the following problems:
\begin{enumerate}
\item
The operation of passing from a pointwise function to the `abstracted{}' version defined on equivalence classes is done using type-theoretical \verb|Abs_| and \verb|Rep_|.
In paper-based mathematics, on the other hand, everything is done using set theory; hence, a mathematician would probably not know how to use this implementation without first getting some knowledge of the underlying type-theoretical foundations.
\item
This operation must be performed `manually{}' in each separate case: there is no generic definition to do that given a pair $\left( f, R \right)$, where $f$ is the function to be abstracted and $R$ is an equivalence relation on its domain.
In contrast, in our treatment we introduce a function called \verb|quotient| doing exactly this.
\item
$f$ and $R$ are typed as a lambda function and a set-theoretical relation, respectively,\rnote{MK: put respectively at the end of the construct} while there is no reason to preemptively preventing $f$ from being a generic relation (i.e., not necessarily being right unique).
\end{enumerate}


For these reasons, we coded a purely set-theoretical \I{} implementation of this machinery, via three simple definitions and one theorem establishing the right-uniqueness of the abstracted function given basic requirements on the pointwise function.

The first definition gives a map to pass from a point of the domain of a relation \verb|R| to the corresponding equivalence class:
\begin{mytable}
\begin{alltt}
definition projector where 
"projector R = \{(x,R``\{x\}) | x. x ∈ Domain R \}"
\end{alltt}
\end{mytable}
The second definition builds, given a pointwise relation \verb|R| and two equivalence relations \verb|P|, \verb|Q| (working on its domain and codomain, respectively) the corresponding, abstracted relation on the resulting equivalence classes:
\begin{mytable}
\begin{alltt}
definition quotient where "quotient R P Q = 
\{(p,q)| p q. q ∈ (Range (projector Q)) ∧ 
p ∈ Range (projector P) ∧ p × q ∩ R ≠ \{\} \}".
\end{alltt}
\end{mytable}
While this definition is typically given for a function \verb|R| and equivalence relations \verb|P| and \verb|Q|, it still makes sense if these additional conditions are not satisfied.
This allows us to lift these requirements in some preparatory lemmas before assuming them in the following main result:
\label{RefWellDefinednessTheorem}
\begin{mytable}
\begin{alltt}
lemma l23: assumes "compatible f P Q" "runiq f" "trans P" "sym P"
"equiv (Domain Q) Q" shows "runiq (quotient f P Q)",
\end{alltt}
\end{mytable}
where the predicates \verb|equiv X P|, \verb|trans P|, \verb|sym P| exist in the \I{} library, and state, respectively, that \verb|P| is an equivalence relation over the set \verb|X|; that \verb|P| is a transitive relation; and that \verb|P| is a symmetric relation.

Note that in \I{} there is a definition for a quotient of a relation \verb|R| written as \verb|quotient R| as the set of all equivalence classes associated with \verb|R|. Here, however, we assume a  function (or relation) \verb|f| with respect to relations \verb|P| and \verb|Q| and define the quotient of \verb|f| with respect to \verb|R| as a function with the domain \verb|quotient R|.

The notion of compatibility above asks that \verb|f| respects \verb|P| and \verb|Q|:\smallskip

\noindent\begin{minipage}{0.6\textwidth}
\begin{mytable}
\begin{alltt}
definition compatible where  
"compatible R P Q =
 (∀ x . (R``(P``\{x\}) ⊆ Q``(R``\{x\})))"
\end{alltt}
\end{mytable}

Note that the definition of \verb|compatible| is a generalization of the usual commutativity of the application of functions as displayed to the right.
\end{minipage}\qquad
\begin{minipage}{0.35\textwidth}
\includegraphics[width=1\textwidth]{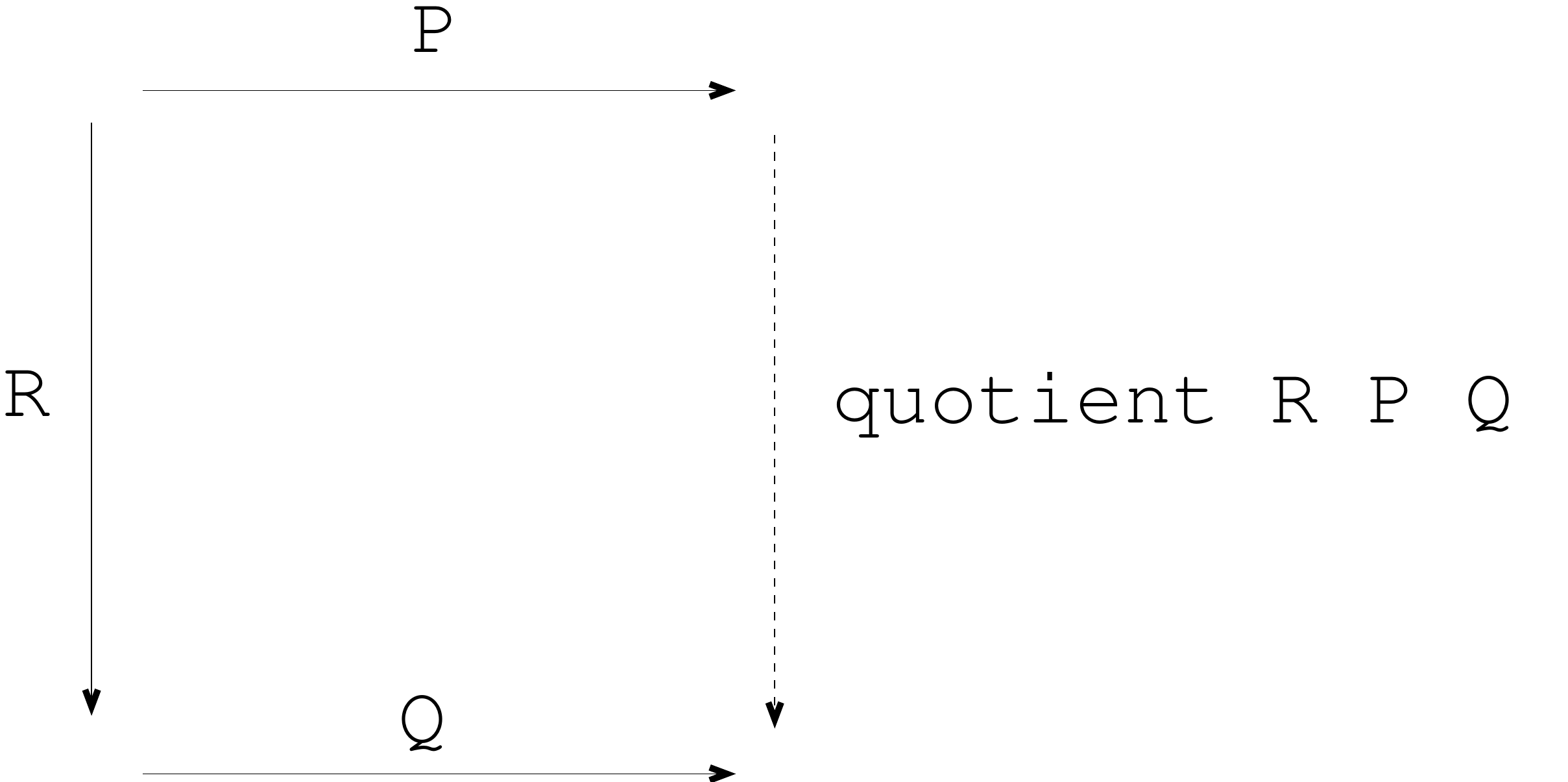}
\end{minipage}\bigskip\rnote{MK:Changed wrapfigure to minipage}

It should be noted that 
this condition is not required at the time of defining a quotient, which can be defined for any triple of relations;  
rather, it is only required when asking that the quotient behaves in the expected way, as stated by \verb|lemma l23| above.
This allows us to freely use the construction \verb|quotient| in advance, to show a number of intermediate results not requiring compatibility themselves. For example, the following lemma holds for general relations:


\begin{small}
\begin{alltt}
lemma quotientFactors: assumes "equiv (Domain p) p" "equiv (Domain q) q"
shows "quotient r p q = (projector p)¯ O r O (projector q)".
\end{alltt}
\end{small}
where \verb|R¯| is the converse of the relation \verb|R| and \verb|O| stands for relation composition.

We also note that, for similar reasons, we devised a definition of compatibility taking any triple of relations as arguments: as in the definition of quotient, one can use \verb|compatible R P Q| even before showing that \verb|R| is right-unique and that \verb|P|, \verb|Q| are equivalence relations.
In the latter case, however, the definition of compatibility reduces to asking that any two \verb|Q|-equivalent points have \verb|P|-equivalent images through \verb|R|.


\section{Injective functions and partitions}
\label{RefDualDefs}
In section \ref{RefFuncAsGraphApplication}, we introduced the mathematical description of a single-good auction.
An important family of more complex schemes is given by \emph{combinatorial} auctions.
In these, there are several objects at stake (a set of goods $G$), and each bidder (of a set $N$) can bid for each possible combination of them.
The outcome of the auction is still described by a pair of maps $\left( a, p \right)$, yielding respectively what a bidder gets and how much she has to pay.
However, now the ``what a bidder gets{}'' part must be represented by a mathematical object more articulate than a $\left\{ 0,1 \right\}$-valued function.
It is represented by a partition of $G$, and by an injective function (or injection) from that partition to $N$.
Treating the latter function inside set-theory gives an advantage: the pair (partition, injection) is conveniently represented by the injection alone, because the partition of $G$ will be simply the domain of the former.
This allowed us to type the relevant objects plainly as follows:
\begin{mytable}
\begin{alltt}
type_synonym bidder = "nat"
type_synonym goods = "nat set"
type_synonym allocation_rel = "(goods × bidder) set".
\end{alltt}
\end{mytable}

Since we wanted to extract code from our \I{} formalization, we had to implement recursive definitions for both the set of all possible partitions of a finite set and for the set of all possible injections from a finite set to another finite set.
In both cases, those recursive definitions turned out to be inconvenient when it came to prove mathematical facts involving them. Hence a separate, more natural definition was needed that is equivalent but not computable.

We illustrate this dual approach in the case of injections.
\begin{mytable}
\begin{alltt}
fun injections_alg :: 
"'a list ⇒ 'b::linorder set ⇒ ('a × 'b) set list" where 
"injections_alg [] Y = [\{\}]" |
"injections_alg (x # xs) Y = concat 
[[R+*\{(x,y)\}. y←sorted_list_of_set (Y-Range R)]. 
                          R ← injections_alg xs Y]"
\end{alltt}
\mycaption{Constructive definition of injection}\label{table:injective_alg}
\end{mytable}

In Table~\ref{table:injective_alg}, we find a definition of all injective functions between two finite sets \verb|X| and \verb|Y|, which recurs on \verb|X|, while in Table~\ref{table:injections} the set of all injective functions is defined axiomatically.
\begin{mytable}
\begin{alltt}
definition injections :: "'a set ⇒ 'b set ⇒ ('a × 'b) set set"
where "injections X Y = 
\{R. Domain R=X ∧ Range R⊆Y ∧ runiq R ∧ runiq(R¯)\}"
\end{alltt}
\mycaption{Axiomatic definition of injection}\label{table:injections}
\end{mytable}

We use the constructive definition in computations and the axiomatic for proofs and mathematical manipulations. 
In order to do that we have to prove their equivalence in the theorem stated in Table~\ref{table:theoremInjections_equiv}.
\begin{mytable}
\begin{alltt}
theorem injections_equiv:
fixes xs::"'a list"
and Y::"'b::linorder set"
assumes non_empty: "card Y > 0"
shows "distinct xs ⟹ 
(set(injections_alg xs Y)::('a×'b)set set)=injections (set xs) Y"
\end{alltt}
\mycaption{\Junction{} of constructive and axiomatic definitions of injection}\label{table:theoremInjections_equiv}
\end{mytable}
Similarly, we have a constructive and an axiomatic definition for partitions (see Tables~\ref{table:partition_alg} and \ref{table:partition_ax}, respectively).
\begin{mytable}
\begin{alltt}
definition insert_into_member_list
:: "'a ⇒ 'a set list ⇒ 'a set ⇒ 'a set list"
where "insert_into_member_list new_el Sets S = 
	(S ∪ \{ new_el \}) # (remove1 S Sets)"

definition coarser_partitions_with_list 
::"'a ⇒ 'a set list ⇒ 'a set list list" where 
"coarser_partitions_with_list new_el P = (\{ new_el \} # P)
  #
  (map ((insert_into_member_list new_el P)) P)"

definition all_coarser_partitions_with_list 
::"'a ⇒ 'a set list list ⇒ 'a set list list" where 
"all_coarser_partitions_with_list elem Ps = 
	concat (map (coarser_partitions_with_list elem) Ps)"

fun all_partitions_list :: "'a list ⇒ 'a set list list" where 
"all_partitions_list [] = [[]]" |
"all_partitions_list (e # X) = 
	all_coarser_partitions_with_list e (all_partitions_list X)"
\end{alltt}
\mycaption{Constructive definition of partition}\label{table:partition_alg}
\end{mytable}
The constructive definition above represents a partition of a finite set as a list of (disjoint) subsets of it; it works by induction on the cardinality of the set to be partitioned, as follows.
Given a set of cardinality $n+1$, we write it as a disjoint union $X \cup \left\{ x \right\}$, and, given a partition $P$ of $X$, we insert $x$ into each set belonging to $P$ (this is done by the operator \verb|insert_into_member_list| above).
We thus obtain $|P|$ many partitions of $X \cup \left\{ x \right\}$, to which we add the distinct partition $\left\{ \left\{ x \right\} \right\} \cup P$.
Making $P$ range over all possible partitions of $X$, we obtain in this way all possible partitions of $X \cup \left\{ x \right\}$.

When implementing this idea into the code above, however, we chose to represent the finite set $X$ as a list (rather than a set) of elements of it, and a partition of it as a list (rather than a set) of subsets of it.
This is because the algorithm just described is iterated in two ways: first, $x$ is inserted into each set of a partition; secondly, the construction is iterated for each possible partition of $X$.
Since iterations are easier with lists than with finite sets \cite{nipkow2005proof}, we adopted that choice.

\begin{mytable}
\begin{alltt}
definition is_partition where
"is_partition P = (∀ X∈P . ∀ Y∈ P . (X ∩ Y ≠ \{\} ↔ X = Y))"

definition is_partition_of (infix "partitions" 75)
where "is_partition_of P A = (\(\bigcup\) P = A ∧ is_partition P)"

definition all_partitions where 
"all_partitions A = \{P . P partitions A\}."
\end{alltt}
\mycaption{Axiomatic definition of partition}\label{table:partition_ax}
\end{mytable}

It should be noted that passing from a list to a finite set (via the operator \verb|set|) is easier than the converse, hence the equivalence theorem for partitions in Table~\ref{table:theoremPartition_equiv} is stated taking as input data a list \verb|xs|:
\begin{mytable}
\begin{alltt}
theorem all_partitions_paper_equiv_alg:
  fixes xs::"'a list"
  shows "distinct xs ⟹ 
  set(map set (all_partitions_list xs)) = all_partitions(set xs)"
\end{alltt}
\mycaption{\Junction{} of constructive and axiomatic definitions of partition}\label{table:theoremPartition_equiv}
\end{mytable}

Thanks to the constructive definitions above, we were able to 
extract, from \I{} code \cite{haftmann2010code}, executable code running Vickrey combinatorial auctions
\cite{CKLR:SoundCombVickCode13}.
Furthermore, we were able to prove fundamental theoretical properties about those auctions by using the non-constructive versions of those definitions; for example, that the price for any bidder is non-negative.
The \junction{} theorems in Tables~\ref{table:theoremInjections_equiv} and \ref{table:theoremPartition_equiv}
allow to certify that such theoretical properties hold also for the extracted code.

\lnote{MC: Added for clarification following reviewer2.}
We note that, while non-constructive versions hold in the general case, constructive versions are obviously limited to the finite case (e.g., calculating all the partitions of a finite set).
Therefore, the two \junction{} theorems must restrict to the finite case: this is reflected by the fact that the argument of \verb|injections| and \verb|all_partitions|, appearing in Tables~\ref{table:theoremInjections_equiv} and \ref{table:theoremPartition_equiv}, respectively, is \verb|set xs|.\rnote{MK: respectively after}
This means that such an argument is automatically a finite set, being the result of converting a list (\verb|xs|) to a set using the \I{} function \verb|set|.
We stress the fact that our approach allows to prove most theorems 
(e.g., thesis~\eqref{RefMaskin2Thesis} at the end of section~\ref{RefFuncAsGraph}) without restricting to the finite case, and to add the additional hypothesis of finiteness only and exactly for the theorems needing it.

\section{Application of quotients 
to auctions}
\label{RefApplication}



We now give enough formalization details to illustrate the exact point in our proofs of auction theory in which we needed to employ \verb|projector| and \verb|quotient|, introduced in section \ref{RefQuotients}.

To prove the thesis \eqref{RefMaskin2Thesis} at the end of section~\ref{RefFuncAsGraph}, we needed to explicitly build the \verb|t| appearing in the definition of \verb|genvick|. 
This function is uniquely determined by \verb|a| and \verb|p|; however, the latter takes \verb|b| as an argument, while \verb|t| takes \verb|b--i| (which represents a bid vector with bidder \verb|i|'s bid removed, and is called a \emph{reduced bid}).
The algebraic way to pass from \verb|b| to \verb|b--i| is to consider an equivalence relation associating any two \verb|b| \verb|b'| differing at most in the point \verb|i|. Correspondingly, to obtain from \verb|p| a function on arguments of the form \verb|b--i|, we form the quotient of \verb|p| according to that equivalence relation.
The equivalence relation we need is exactly the kernel%
\footnote{We recall that the kernel of a function $f$ is the equivalence relation $\circ_f$ given by $x_1 \circ_f x_2 \iff{} f \left( x_1 \right) = f \left( x_2 \right)$. See \cite[Definition~1.18]{bergman2011universal}. The kernel notion was missing in the \I{} library, and we also provided it in ours.}
of the following function:
\begin{mytable}
\begin{alltt}
definition reducedbid:: "bidder ⇒ (bid × allocation) set ⇒ 
(bid × bidder set × bid × allocation) set"
where "reducedbid i a = 
\{(b, (Domain b, b outside \{i\}, a ,, b))| b. b ∈ Domain a\}".
\end{alltt}
\end{mytable}
So that the explicit construction of \verb|t|, used inside the proof for \eqref{RefMaskin2Thesis}, ends up as
\begin{mytable}
\begin{alltt}
"λx. reducedprice p i a ,, (\{i\} ∪ Domain x, x, Min (Range a))", 
\end{alltt}%
\end{mytable}
where
\begin{mytable}
\begin{alltt}
definition reducedprice:: "(bid × price) set ⇒ bidder ⇒ 
(bid × allocation ) set ⇒ 
((bidder set × bid × allocation) × price) set" 
where "reducedprice p i a = 
(projector ((reducedbid i a)¯)) O 
(quotient p (Kernel (reducedbid i a)) Id) O 
((projector Id)¯)",
\end{alltt}
\end{mytable}
\verb|Id| being the identity function. 
With this definition in place, an important part of the theorem consists in showing that \verb|reducedprice| is right-unique.

This reduces to showing that 
\verb|quotient p |
\verb|(Kernel(reducedbid i a))|
\verb| Id| appearing in the definition above is right-unique. 
Thanks to \verb|lm23| (see section \ref{RefWellDefinednessTheorem}), this in turn means proving the compatibility between \verb|p| and the equivalence relation just introduced, i.e., \verb|Kernel (reducedbid i a)|, 
which is provided by 
\begin{mytable}
\begin{alltt}
lemma l24b: assumes "functional (Domain a)" "Domain a ⊆ Domain p" 
"dom4 i a p" "runiq p" shows 
"compatible p (Kernel (reducedbid i a)) Id".
\end{alltt}
\end{mytable}

\section{Discussion and related work}
\label{RefSetTheory}
\label{RefRelatedWork}
The work presented here is based on the specific set theory implemented in \I{}/HOL, sometimes called simply-typed set theory \cite[Section~6.1]{paulson2013proof}, \cite[Section~1]{Paulson:stfv93}%
.
It differs from standard set theories, as Zermelo-Fraenkel (ZF), in that the primitives of the latter are encoded using primitives of \hol{}, as follows: a given set $p$ is actually a term $p$ of type $\tau \Rightarrow \text{bool}$, and the writing $x \in p$ is actually the application $p\ x$.
This means that in \I/\Hol{} one cannot write a set like $ \left\{ x, \left\{ x \right\} \right\}$ (which is not well-typed): the standard hierarchy of ZF is no longer constructable.
As long as the objective is to formalize ordinary 
(as defined in \cite[Section~I.1]{simpson2009subsystems}) mathematics, this usually causes no problem:
while this difference prevents the encoding of relevant mathematical objects (e.g., natural numbers, integers, reals, cartesian products) as usually done in ZF, those objects can be directly represented in \hol{}.
This means that, for many mathematical branches, what is lost with respect to ZF is limited to its more technical side-effects, as $\pi \cap \mathbb{Q} \neq \emptyset$, or $1 \cap 2 \subseteq 3$.
Since the latter are often regarded as strange or meaningless writings \cite{leinster2012rethinking}, \cite[Section~1]{Paulson:stfv93}, this could even be a desirable consequence.

On the other hand, when trying to exploit technical ZF `hacks{}', as we tried to do with \verb|eval_rel2|, problems can arise: we discussed that in section \ref{RefEvalRel2}.
Moreover, ZF is a more appropriate tool for studying the remaining branches of mathematics, starting with set theory itself.

The Z pattern catalogue \cite[Section~IV]{valentine2004az} allows to represent functions as sets of pairs, as done here.
However, Z is a specification language, while we are interested in both specification and implementation.
\M{} is based on untyped set theory, thus modelling functions and relations as done here, and also provides many relevant existing theorems in its library; however, it is not possible to extract code or to do computations in general.
The Ssreflect extension library for the Coq proof assistant is extensive and provides a lot of operations; on the other hand, most of this material seems to apply to functions as represented by lists, thereby inherently limiting to the finite case.
More details on how \I{}/HOL generally compares with other systems are 
in the comparative study \cite{la-ca-ke-mo-ro-we-ww-13}.

\section{Conclusions}\label{RefConclusions}

We have built an extensive library of results about functions, represented as right-unique relations as an alternative to the lambda abstractions that have so far been typical of \I{}/HOL.
In this paper we explained how we employed this alternative technique, and the concepts in our library, in the application domain of auction theory.
Our library ranges from simpler constructions such as pasting a relation onto another, to more sophisticated ones such as quotients.\rnote{MK: I think, we should mention again the one advantage mentioned previously, namely, that set theory allows for the generalization of theorems.}

We described how set-theoretical constructs can concisely express constructions that frequently occur in formalization, giving concrete examples from the application domain of auction theory.
In the particular case of representing functions, we discussed the advantages and disadvantages of the set-theoretical representations with respect to the alternative, more natural (given the foundations of \I/HOL) lambda representation.
We also showed that computability, a typical feature of lambda functions, can be achieved for set-theoretical functions in \I/HOL.
Moreover, even in those cases in which it cannot be achieved, we showed how we were still able to employ a dual approach, giving non-constructive, more expressive definitions along with constructive (and thus computable) ones, finally proving their equivalence through \junction{} theorems.
This allows us to obtain the best of both worlds, that is, expressiveness and proof-friendly definitions together with computability, at the cost of having to prove the additional \junction{} theorems.




Finally, we took our application of this approach to a more advanced, abstract level by describing in a simple, close-to-paper set-theoretical style notions such as \emph{quotients}, \emph{compatibility} and \emph{kernel}.
We presented a concrete application of this material in a hands-on case encountered in formalizing auction theory within the \fo{} project.
We think that our library is rich and generic enough to be of possible use to other \I{} users, and we hope that this paper can serve as a first guiding example to show how it can be employed.


\begin{spacing}{0.95}
\printbibliography
\end{spacing}
\end{document}